\documentclass[12pt]{article}
\usepackage{cite}
\usepackage{a4,amsmath}
\usepackage{float}

\renewcommand{\d}{\mathrm{d}}
\renewcommand{\i}{\mathrm{i}}
\newcommand{\fl}{}

\newcommand {\BR}{\nonumber\\}
\newcommand {\be}{\begin{equation}}
\newcommand {\ee}{\end{equation}}
\newcommand {\bea}{\begin{eqnarray}}
\newcommand {\eea}{\end{eqnarray}}
\newcommand{\INDENT}{\quad\quad}

\begin{document}
\title { The logarithmic perturbation theory for bound
states in spherical-symmetric potentials via the $\hbar$-expansions
}
\author{I. V. Dobrovolska \thanks{dobrovolska@resonance.zp.ua},
R. S. Tutik \thanks{tutik@ff.dsu.dp.ua}\\
Theoretical Physics Department \\ Dniepropetrovsk National
University \\
13 Naukova str., Dniepropetrovsk 49050, Ukraine }

\maketitle

\begin{abstract}
The explicit semiclassical treatment of the logarithmic perturbation
theory for the bound-state problem for the spherical anharmonic
oscillator and the screened Coulomb potential is developed. Based
upon the $\hbar$-expansions and suitable quantization conditions a
new procedure for deriving perturbation expansions is offered.
Avoiding disadvantages of the standard approach, new handy recursion
formulae with the same simple form both for ground and excited
states have been obtained. As examples, the perturbation expansions
for the energy eigenvalues of the quartic anharmonic oscillator and
the Debye potential are considered.
\end{abstract}

\section{Introduction}

The main task in application of the quantum mechanics is to solve
the Schr\"odinger equations with different potentials.
Unfortunately, realistic physical problems can practically never be
solved exactly. Then one has to resort to some approximations. Most
widely used among them is the perturbation theory. However, the
explicit calculation with the Rayleigh -- Schr\"odinger perturbation
theory, described in most quantum mechanics textbooks, runs into the
difficulty of the summation over all intermediate unperturbed
eigenstates. To avoid this difficulty, various alternative
perturbation procedures have been proposed \cite{b1, b2, b3, b4, b5,
b6, b7, b8, b9, b10, b11, b12, b13}.

Nevertheless up to now, one of the principal approximation
techniques is the logarithmic perturbation theory \cite{b14, b15,
b16, b17, b18, b19, b20, b21}. Within the framework of this
approach, the conventional way to solve a quantum-mechanical
bound-state problem consists in changing from the wave function to
its logarithmic derivative and converting the time-independent
Schr\"odinger equation into the nonlinear Riccati equation.

In the case of ground states, the consequent expansion in a small
parameter leads to handy recursion relations that permit us to
derive easily the corrections to the energy as well as to the wave
function for each order. However, when radially excited states are
considered, the standard technique of the logarithmic perturbation
theory becomes extremely cumbersome and, practically, inapplicable
for describing higher orders of expansions.

Recently, a new procedure based on specific quantization conditions
has been proposed to get series of the logarithmic perturbation
theory via the $\hbar$-expansion technique within the framework of
the one-dimensional Schr\"odinger equation \cite{b22}. Avoiding the
disadvantages of the standard approach, this straightforward
semiclassical procedure results in new handy recursion formulae with
the same simple form both for the ground state and excited states.

The object of the present work is to extend the above mentioned
formalism to the bound-state problems within the framework of the
three-dimensional Schr\"odinger equation with  central potentials,
such as the anharmonic scillator potential and the screened Coulomb
one, which are widely used in practice.

\section{Basic concepts of the method}

We study the bound-state problem for a non-relativistic particle
moving in a central potential admitted bounded eigenfunctions and
having in consequence a discrete energy spectrum. Let us therefore
consider the reduced radial part of the Schr\"{o}dinger equation
\begin{equation}
 -\frac{{\hbar}^2}{
{2m}}U''(r)+\left({\frac{{\hbar}^2l(l+1)}{2mr^2}+V(r)}\right)U(r)=EU(r),
\label{1}
\end{equation}
with the effective potential having  only  one simple minimum.

 Following usual practice, we apply the substitution, $ C(r)=
\hbar U'(r) / U(r) $, accepted in the logarithmic perturbation
theory and go over from the Schr\"{o}dinger equation (\ref{1}) to
the Riccati equation
\begin{equation}
\hbar C'(r)+C^{2}(r)=\frac{{\hbar}^2l(l+1)}{r^2}+2mV(r)-2m E.
\label{3}
\end{equation}

According to our assumption, we are seeking the eigenvalues and the
eigenfunctions of this equation  explicitly in a semiclassical
manner with series expansions in powers of the Planck constant
\begin{equation}
E=\sum^{\infty}_{k=k1}{E_{k}\hbar^{k}},\;\;\;\;\;
C(r)=\sum^{\infty}_{k=k2}C_{k}(r)\hbar^{k}, \label{4}
\end{equation}
where the order in $\hbar$ of these quantities, i.e. the values of
$ k1 $ and $ k2 $, should be defined as a preliminary.

As in the standard approach, the substitution of these expansions
into the Riccati equation leads to the simple recursion system. This
system can be solved successively in the case of ground states,
while the description of the excited states has some problems with
taking into account the nodes of wave functions. For avoiding these
problems, we shall attempt to use the quantization condition and the
formalism of the theory of functions of a complex variable.

Remind that, in the complex plane, a number of zeros $N$ of a
regular function inside a closed contour is defined by the principle
of argument known from the  complex analysis. Being applied to the
logarithmic derivative, $C(r)$, it means that
\begin{equation}
\frac{1}{2\pi\i}\oint{C(r)\,\d{r}}=
\frac{1}{2\pi\i}\sum^{\infty}_{k=0}{\hbar^{k}\oint{C_{k}(r )\,\d
r}}=\hbar N. \label{5}
\end{equation}

This quantization condition is exact and is widely used for deriving
higher-order corrections to the WKB-approximation \cite{b23,b24} and
the $1/N$-expansions \cite{b25,b26,b27}. There is, however, one
important point to note. Because the radial and orbital quantum
numbers, $n$ and $l$, correspondingly, are specific quantum notions,
the quantization condition (\ref{5}) must be supplemented with a
rule of achieving a classical limit for these quantities. It is this
rule that stipulates the kind of the semiclassical approximation.

In particular, within the framework of the WKB-approach the
passage to the classical limit is implemented using the rule
\begin{equation}
\hbar\to 0,\;\;n \to\infty,\;\; l\to\infty,\;\;\hbar n={\rm
const},\;\;\hbar l={\rm const},\label{6}
\end{equation}
whereas the $1/N$-expansion requires the condition
\cite{b25,b26,b27}
\begin{equation}
\hbar\to 0,\;\; n={\rm const},\;\; l\to\infty,\;\;\hbar n\to
0,\;\;\hbar l={\rm const}.\label{7}
\end{equation}

The proposed semiclassical treatment of the logarithmic
perturbation theory involves the alternative possibility:
\begin{equation}
\hbar\to 0,\;\; n={\rm const},\;\;l={\rm const},\;\; \hbar n\to
0,\;\;\hbar l\to 0. \label{8}
\end{equation}

With the last rule, the right-hand side of the equation (4) has the
first order in $\hbar$ and the quantization condition  now takes the
simple form
\begin{equation}
\frac{1}{2\pi\i}\oint{C_1(r)\, \d r}=N,\;\;\;\;\;\;\;\;
\frac{1}{2\pi\i}\oint{C_{k}(r )\,\d r}=0, \quad \;k\neq1.
\label{10}
\end{equation}

However, this definition of the quantization condition is incomplete
since we have not pointed out the path of integration. We shall now
show that the suitable  choice of the contour of integration and the
consequent integration with application of the Cauchy residue
theorem   easily solves the problem of  describing   radially
excited states.

\section{The anharmonic oscillator}
{\it Quantization conditions.} The discussion of details of the
proposed technique we begin with the case of the anharmonic
oscillator potential which is given by a symmetric function $V(r)$
that can be written as a Taylor series expansion
\begin{equation}
V(r)=\frac{1}{2}m\omega^{2}r^{2}+\sum_{i{\geq}1}f_{i}r^{2i+2}.
\label{2}
\end{equation}

In the first place, let us consider the rule (7) of achieving the
classical limit from the physical point of view. Since $\hbar l
\rightarrow 0$ as $\hbar \rightarrow 0$, the centrifugal term,
$\hbar^2 l \left(l+1\right)/r^2$, has the second order in $\hbar$
and disappears in the classical limit that corresponds to falling a
particle into the center. This means that a particle drops into the
bottom of the potential well as $\hbar \rightarrow 0$ and its
classical energy becomes $E_0 = \min V(r) = 0$. Hence, the series
expansions in powers of the Planck constant for the energy
eigenvalues and the $ C(r)$ must now read as $E =
\sum_{k=1}^{\infty}{E_k\hbar^k}$ and  $ C(r)=
\sum_{k=0}^{\infty}C_{k}(r)\hbar^k$.

Upon inserting  these expansions  into the Riccati equation
(\ref{3})  and collecting coefficients of equal powers of $\hbar$,
we obtain the following hierarchy of equations
\begin{eqnarray}
C_{0}^{2}(r)&=&2 \; m V(r),\nonumber\\
C'_{0}(r)+2 \; C_{0}(r)C_{1}(r)&=&-2 \; m E_{1},\nonumber\\
C'_{1}(r)+2 \; C_{0}(r)C_{2}(r)+C_{1}^{2}(r)&=&\frac{l(l+1)}{r^2}-2 \; m E_{2},\nonumber\\
&\cdots& \nonumber\\
C'_{k-1}(r)+\sum_{i=0}^{k}C_{i}(r)C_{k-i}(r)&=&-2 \; m
E_{k},\;\;k>2. \label{9}
\end{eqnarray}

In the case of ground states,  this recurrence system can be solved
as straightforwardly as in the standard approach. For excited
states, we intend to  take into account the nodes of the wave
function with the quantization condition (8) for which we must
define the contour of integration.

It should be stressed that our technique is quite distinguished from
the WKB method not only in the rule of achieving a classical limit
but also in the choice of a contour of integration in the complex
plane. With a view to elucidate the last difference let us now
sketch out the WKB treatment of this bound-state problem. In the
complex plane, because the potential is described by the symmetric
function (\ref{2}), there are two pairs of turning points, i.e.
zeros of the classical momentum, on the real axis. Therefore we have
two cuts between these points: in the region $r>0$ as well as in the
region $r<0$. In spite of only one cut lies in the physical region
$r>0$, the contour of integration in the WKB quantization condition
has to encircle both cuts for the correct result for the harmonic
oscillator to be obtained \cite{b28}.

In our approach, when a particle is dropping into the bottom of the
potential well these four turning points are drawing nearer and, at
last, are joining together at the origin. Consequently, all zeros of
the wave function are now removed from both positive and negative
sides of the real axis into the origin and our contour of
integration must enclose only this point and no other singularities.

Further, let us count the multiplicity of a zero formed in the wave
function  at $r = 0$. Imposed by the requirement of the regularity,
the behavior $r^{l+1}$ as $r \rightarrow 0 $ brings the value $l+1$.
The number of nodes  in the physical region $ r > 0$ is equal to the
radial quantum number $n$. But, because the potential (\ref{2}) is a
symmetric function, the same number of zeros must be in the region
$r<0$, too. Then the total number of zeros inside the contour
becomes equal to $N=2n+l+1$.

For  evaluation of the contour integrals in the quantization
condition (8), let us consider the system (\ref{9}) and investigate
the behavior of the functions $C_k(r)$. From the first equation of
this system, it follows instantly that the $C_0(r)$ can be written
as
\begin{equation} \fl
C_0(r) = - \left[2 \, m\,  V(r) \right]^{1/2} = -m\, \omega\, r
\left( 1 + \frac{2}{m\, \omega^2} \sum_{i=1}^{\infty}{f_i
\,r^{2i}}\right)^{1/2} =\, r \sum_{i=0}^{\infty}{C_i^0 \,r^{2i}},
\label{11}
\end{equation}
where the minus sign is chosen from the boundary conditions and
coefficients $C_i^0$ are defined by parameters of the potential
through the relations
\begin{equation}
C^{0}_{0}=-m\omega,\;\;\;\;\;\;\;\; C^{0}_{i}=\frac{1}{2m\omega}
\left({\sum_{p=1}^{i-1}{C^{0}_{p} C^{0}_{i-p}-2 m f_{i}}}\right),
\;i\geq 1.\label{12}
\end{equation}

At the origin, on account of the equality $C_0(0)=0$, a simple pole
arises for the function $C_1(r)$, while $C_k(r)$ has a pole of the
order $\left(2k-1\right)$. Thus $C_k(r)$ can be represented by the
Laurent series
\begin{equation}
C_{k}(r)= r^{1-
2k}\sum^{\infty}_{i=0}{C^{k}_{i}r^{2i}},\;\;\;\;\;k\geq 1.\label{13}
\end{equation}

Finally, with applying the residue theorem, the quantization
condition (8) expressed   explicitly in terms of the coefficients
$C_i^k$ takes the especially simple form
\begin{equation}
C^{k}_{k-1}= N \delta_{1, k},\label{14}
\end{equation}
where $N=2n+l+1$ .

It is this quantization condition that makes possible the common
consideration of the ground and excited states and permits us to
derive the simple recursion formulae.

{\it Recursion formulae and the example of application.} The
substitution of the series (\ref{12}) and (\ref{13}) into the system
(\ref{9}) in the case $i \neq k-1$ yields the recursion relation for
obtaining the Laurent-series coefficients of the logarithmic
derivative of the wave function
\begin{equation} \fl
C^{k}_{i}=-{\frac1{2C^{0}_{0}}} \left[{(3-2k+2i) C^{k-1}_{i
}+\sum_{j=1}^{k-1}\sum_{p=0}^{i} C^{j}_{p}C^{k-j}_{i-p}
+2\sum_{p=1}^{i}C^{0}_{p}C^{k}_{i-p}}-l(l+1)\delta_{2,k}\delta_{0,i}\right].\label{15}
\end{equation}

If  $i=k-1$, by equating the  expression (\ref{15}) for $C^k_{k-1}$
 to the quantization condition (\ref{14}) we arrive at the
recursion formulae for the energy eigenvalues
\begin{equation}
2mE_{k}=- C^{k-1}_{k-1}- \sum_{j=0}^{k}\sum_{p=0}^{k-1}
C^{j}_{p}C^{k-j}_{k-1-p} \; .\label{16}
\end{equation}

Derived in this way,  first corrections to the energy eigenvalues of
the spherical anharmonic oscillator  take the form
\begin{eqnarray}
&& E_1  =  \frac{1 + 2\,N}{2} \,\omega ,\;\;\;\; E_2 =  \frac{\left(
            3 - 2\,L + 6\,\eta  \right) \,{f_1}}{4\,m^2\,{\omega }^2},
\BR && E_3  =  \frac{1 + 2\,N}{8\,m^4\,{\omega }^5} \,\left(
\left( -21 + 9\,L -
                  17\,\eta  \right) \,{{f_1}}^2 +
            m\,\left(
                15 - 6\,L +
                  10\,\eta  \right) \,{\omega }^2\,{f_2} \right),
\BR &&  E_4 = \frac{1}{16\,m^6\,{\omega }^8}\bigg(\left(
            333 + 11\,L^2 - 3\,L\,\left( 67 + 86\,\eta  \right)  +
              3\,\eta \,\left(
                347 + 125\,\eta  \right)  \right) \,{{f_1}}^3 \BR && \INDENT -
        6\,m\,\left(
            60 + 3\,\left( -13 + L \right) \,L + 175\,\eta  - 42\,L\,\eta  +
              55\,{\eta }^2 \right) \,{\omega }^2\,{f_1}\,{f_2}  \BR && \INDENT
       + m^2\,\left(
            6\,L^2 - 12\,L\,\left( 6 + 5\,\eta  \right)  +
              35\,\left(
                3 + 2\,\eta \,\left(
                    4 + \eta  \right)  \right)  \right) \,{\omega
}^4\,{f_3}\bigg), \BR && E_5 = - \frac{1 + 2\,N
}{128\,m^8\,{\omega }^{11}}\bigg(\left(
                    30885 + 909\,L^2 -
                      27\,L\,\left( 613 + 330\,\eta  \right)  + \eta \,\left(
                        49927 + 10689\,\eta  \right)  \right) \,{{f_1}}^4
                        \BR && \INDENT -
                4\,m\,\left(
                    11220 + 393\,L^2 -
                      6\,L\,\left( 1011 + 475\,\eta  \right)  + \eta \,\left(
                        16342 +
                          3129\,\eta  \right)  \right) \,{\omega \
}^2\,{{f_1}}^2\,{f_2}\BR && \INDENT+
                16\,m^2\,\left(
                    33\,L^2 - L\,\left( 501 + 190\,\eta  \right)  +
                      63\,\left(
                        15 + \eta \,\left(
                            19 + 3\,\eta  \right)  \right)  \right) \,{\omega
}^4\,{f_1}\,{f_3} \BR && \INDENT +
                2\,m^2\,\left(
                        3495 + 138\,L^2 + 4538\,\eta  + 786\,{\eta }^2 -
                          30\,L\,\left(
                            63 + 26\,\eta  \right)  \right) \,{\omega }^4\,{{f_2}}^2  \BR && \INDENT-
                    4\,m^3\,\left(
                        30\,L^2 - 20\,L\,\left( 24 + 7\,\eta  \right)  +
                          63\,\left(
                            15 + 2\,\eta \,\left(
                                8 + \eta  \right)  \right)  \right) \,{\omega}^6\,{f_4} \bigg),\label{17}
\end{eqnarray}
where $N = 2 \; n+ l + 1 $, $\eta = N \left(N + 1 \right)$, $L = l
(l+1)$.

As it was expected, the obtained expansion is indeed the series of
the logarithmic perturbation theory  in powers of the Taylor-series
coefficients for the potential function, with the first
approximation being equal to the energy of the three-dimensional
harmonic oscillator
\begin{equation}
E_1 = \left( 2 n + l + \frac{3}{2} \right) \omega.
\end{equation}

Thus, the problem of obtaining the energy eigenvalues and
eigenfunctions for the bound-state problem for the anharmonic
oscillator  can be considered solved. The equations
(\ref{15})-(\ref{16})   have the same simple form both for the
ground and excited states and define a useful procedure for the
successive calculation of higher orders of expansions of the
logarithmic perturbation theory.

As an example, we examine  eigenenergies for the anharmonic
oscillator with the potential
\begin{equation}
V(r) = m \omega^2 r^2 /2 + \lambda r^4 ,\; \;\;\;\lambda > 0.
\end{equation}

Then the equations (\ref{17}) are rewritten as
\begin{eqnarray}
&&  E_1  =  \left( \frac{1}{2} + N \right) \,\omega, \;\;\;\; E_2 =
\frac{\left( 3 - 2\,L + 6\,\eta  \right) }{4\,m^2\,{\omega
}^2}\,\lambda ,\BR && E_3 =  \frac{-\left( 1 + 2\,N \right) \,\left(
21 - 9\,L + 17\,\eta  \right)
     }{8\,m^4\,{\omega }^5}\,{\lambda }^2, \BR
&& E_4  =  \frac{\left( 333 + 11\,L^2 - 3\,L\,\left( 67 + 86\,\eta
\right)  +
       3\,\eta \,\left( 347 + 125\,\eta  \right)  \right) }{16\,
     m^6\,{\omega }^8}\,{\lambda }^3,\\
&& E_5  =  \frac{- \left( 1 + 2\,N \right) \,
       \left( 30885 + 909\,L^2 - 27\,L\,\left( 613 + 330\,\eta  \right)  +
         \eta \,\left( 49927 + 10689\,\eta  \right)
       \right) }{128\,m^8\,{\omega }^{11}}\, {\lambda }^4 .\nonumber
\end{eqnarray}
We recall that here $N = 2 \; n+ l + 1 $, $\eta = N \left(N + 1
\right)$, $L = l (l+1)$.

It is readily seen that the use of the $\hbar$-expansion
technique does lead to the explicit perturbation expansion in
powers of the small parameter $\lambda$.

In the case of ground states, obtained expansions for the energy
eigenvalues coincide with those listed in Ref.~\cite{b29}. In the
case of excited states, our corrections coincide with corrections up
to the second order which are just only calculated in
Ref.~\cite{b30}.

 As it is known, the expansions for the anharmonic oscillator
are asymptotic and diverge for any finite value of the parameter
$\lambda$ that requires the use of some procedures of improving the
convergence (for references see \cite{b31}). It should be noted,
that the proposed technique is easily adapted to apply any scheme of
the series renormalization  \cite{b32}.

\section{The screened Coulomb potential}
{\it Quantization conditions.} Now let us consider the case of the
screened Coulomb potential which in common practice  has a form
\begin{equation}
V(r)=\frac{1}{r}F(\kappa, r) . \label{2''}
\end{equation}
where $ \kappa $ is a small parameter.

In what following, we do not single out explicitly the screening
parameter, but incorporate it into coefficients $ V_{i}$ of the
Taylor series expansion of this potential
\begin{equation}
V(r)=\frac{1}{r}\sum_{i=0}V_{i}r^{i}. \label{2'}
\end{equation}

Note, that after performing the scale transformation $ r\to \hbar^2
r$ powers of the screening parameter appear in common with powers of
Planck's constant squared. Hence, the perturbation series must be,
as a matter of fact, not only $\kappa$-expansions but also the
semiclassical $\hbar^2$ -expansions, too.

In the classical limit, when a particle falls into the center, its
energy  eventually approaches infinity. Hence, the expansions
(\ref{4}) must be represented as $
E=\hbar^{-2}\sum^{\infty}_{k=0}{E_{k}\hbar^{2k}}$ and $C(r)
=\hbar^{-1}\sum^{\infty}_{k=0}C_{k}(r)\hbar^{k}$ that results in the
recurrent system
\begin{eqnarray}
&& C_0^2(r)=-2m E_0\;, \BR && C_0(r)C_1(r)=m\Bigl[
V(r)-E_1\Bigr]\;,\BR && C_1'(r)+2C_0(r)C_2(r)+ C_1^2(r) =\frac
{l(l+1)}{r^2} -2mE_2\;, \BR &&\cdots  \BR &&
C_{k-1}'(r)+\sum_{j=0}^k C_j(r)C_{k-j}(r) =-2mE_{k}\;,\; k>2.
\label{9'}
\end{eqnarray}
which changes only in the first two equations in comparison with
(10).

 Now,  let us consider the choice of the
contour of  integration in the quantization relation. Since in the
classical limit a particle falls into center, the classical turning
points again draw to the origin and the nodes of the wave function
are joined together at $r=0$. Then, as well as in the anharmonic
oscillator case, the contour of integration  must enclose only the
origin. However, now the nodes of the function come into the origin
only from the positive region of the real axis. Thus, with the
number of nodes and the value $l+1$ included,the total number of
zeros in the quantization condition (\ref{5}) becomes equal to
$N=n+l+1$.

Further, from  (\ref{9'}) it appears that $C_0(r)$ is the constant
and its Taylor-series coefficients are
 \begin{equation}
  C_{0}^{0}=-\sqrt{-2mE_0}\;,\;C_{0}^{i}=0,
  \label{12'}
  \end{equation}
Owing to the Coulomb behavior of the potential at the origin, the
$C_1(r)$ has a simple pole at this point, while the function
$C_k(r)$ has a pole of the order $k$ and may be represented as
 \be
  C_k(r)=r^{-k}\sum_{i=0}^\infty{C^{k}_{i}r^i},
  \label{13'}
  \ee
that leads to the known quantization condition
\begin{equation}
C^{k}_{k-1}= N \delta_{1, k},\label{14'}
\end{equation}
where $N=n+l+1$ .

{\it Recursion formulae and the example of application.} After
substitution (\ref{12'})-(\ref{13'}) into (\ref{9'}), when ($i \neq
k$), we have: \bea &&
C^1_i=\frac{m}{C^0_0}\biggl[V_i-E_1\delta_{i,1}\biggr],\BR &&  C^k_i
= -\frac{1}{2C^0_0} \biggl[
     (i-k+1)C^{k-1}_i+
     \sum^{k-1}_{j=1}  \sum^i_{p=0} C^j_p C^{k-j}_{i-p}\BR
&& \INDENT +2mE_k \,\delta_{i,k}- l ( l + 1 )\,\delta_{i,0}
\delta_{k,2}\, \biggr]\;,\;k>1.
 \label{15'}
\eea

In the case $i=k$, from (\ref{14'}) and (\ref{15'}), we obtain
 \bea &&
E_0=-\frac{mV_0^2}{2N^2}\;,\;\; E_1=V_1\;,\BR &&  E_k =
-\frac{1}{2m} \biggl[C^{k-1}_k+
     \sum^{k-1}_{j=1}  \sum^k_{p=0} C^j_p C^{k-j}_{k-p}
        +2C^0_0 C^{k}_{k}
         \biggr]\;,\;k>1 ,
 \label{16'}
\eea that, through the  the Taylor-series coefficients for the
potential function, is
 \bea
&& E_0=-\frac{mV_0^2}{2N^2}\;,\;\;  E_1=V_1\;,\;\; E_2= \frac{\left(
L - 3\,N^2 \right){V_2}}{2\,m\,{V_0}}\;,\BR && E_3=
\frac{N^2}{2\,m^2\,{{V_0}}^2}\,\left( 1 - 3\,L + 5\,N^2 \right)
{V_3}\;,\BR && E_4=\frac{N^2}{8\,m^3\,{{V_0}}^4}\,\left( \left(
        3\,L^2 - 5\,N^2 - 7\,N^4 \right){{V_2}}^2 + \right.\BR
    &&\INDENT \left. \left( 3\,L\left( 2 -
            L \right)- 5\,N^2\left( 5 - 6\,L \right)  -
            35\,N^4 \right) {V_0}\,{V_4} \right)\;,\BR
&& E_5=\frac{N^4}{8\,m^4\,{{V_0}}^5}\,\left( \left( -5\,L\,\left( 2 + 3\,L \right)  + 7\,N^2\left( 9 - 2\,L \right) +
            45\,N^4 \right) \,{V_2}\,{V_3} +\right.\BR
    &&\INDENT \left. \left(12 - 50\,L + 15\,L^2 + 35\,N^2\left( 3 - 2\,L \right) +
        63\,N^4 \right) \,{V_0}\,{V_5} \right),\label{17'}
\eea
 where $N = n+ l + 1 $ and $L = l (l+1)$.

We see that the zero approximation gives  the exact solution for the
Coulomb problem.

As an example of application, we consider the case of the Debye
potential, which is widely used in many branches of physics:
 \be
V(r)=-\frac{\alpha}{r}{\rm{exp}}(-\kappa r).
  \ee

For this potential, the first corrections to the energy eigenvalues
take the form \bea && E_0=-\frac{m\alpha^2}{2N^2}\;,\;\; E_1=\alpha
\,\kappa \;,\;\; E_2= \frac{\left( L - 3\,N^2
\right)}{4\,m}\,{\kappa^2}\;,\BR && E_3=
\frac{N^2}{12\,m^2\,{{\alpha}}}\,\left( 1 - 3\,L + 5\,N^2 \right)
{\kappa^3}\;,\BR && E_4=\frac{ N^2}{192\,m^3\,{\alpha }^2}\,\left(
3\,L\,\left( 2 + 5\,L \right)  -
        5\,\left(11-6\,L \right) \,N^2 + 77\,N^4 \right)
        \,{\kappa }^4\;,\BR
&& E_5=\frac{N^4}{320\,m^4\,{\alpha }^3}\,\left( 4 - 50\,L - 45\,L^2 +
      35\,\left( 7-2\,L \right) \,N^2 + 171\,N^4 \right) \,{\kappa }^5,
\eea
where $N = n+ l + 1 $, and $L = l (l+1)$.

And again we recognize the explicit perturbation expansion in powers
of the small parameter $\kappa$.

Typical results of the calculation with these formulae are presented
in the Table I where the sequences of the partial sums of $K$
corrections to the energy eigenvalues  for the Debye potential $
V(r)=-\alpha \,{\rm{exp}}(-\kappa r)/r$ is compared with  results of
the numerical integration, $E_\text{num}$,   in Coulomb units
$\hbar=m=\alpha=1$. It is seen that for small values of the
screening parameter, the convergence of the series is quite
sufficient for the use them without any renormalization. However,the
results become gradually worse when the screening parameter
increase, as it was pointed for such potentials in
[33].

\begin{table}[ht]\label{table1}
\begin {center}
\begin{tabular}{c|c|c|c}
\hline \vphantom{\Big(} $K$ &  $\,n=0, \;l=0,\;\kappa=0.2\,$ &  $ \,n=1,\;
l=0,\;\kappa=0.04\,$
& $\, n=1,\; l=1,\;\kappa=0.02\,$ \\
\hline
0 & 0.5000000000 & 0.1250000000 & 0.05555555556  \\
1 & 0.3000000000 & 0.0750000000 & 0.03555555556 \\
2 & 0.3300000000 & 0.0825000000 & 0.03805555556  \\
3 & 0.3260000000 & 0.0816250000 & 0.03781555556 \\
4 & 0.3271000000 & 0.0818140625 & 0.03786145556 \\
5 & 0.3266800000 & 0.0817559375 & 0.03784969436  \\
10 & 0.3268179839 & 0.0817715528 & 0.03785241171  \\
15 & 0.3268059572 & 0.0817711705 & 0.03785238868  \\
20 & 0.3268100537 & 0.0817711991 & 0.03785238922 \\
25 & 0.3268067333 & 0.0817711951 & 0.03785238920 \\
\hline
$E_\text{num}$ & 0.3268085112 & 0.0817711958 & 0.03785238920\\
\hline
\end{tabular}
\end{center}
\end{table}

\section{Summary}

In conclusion, a new useful technique for deriving results of the
logarithmic perturbation theory has been developed. Based upon the
$\hbar$-expansions and suitable quantization conditions, new handy
recursion relations for solving the bound-state problem for a
spherical anharmonic oscillator and a static screened-Coulomb
potential have been obtained. These relations can be applied to
excited states exactly in the same manner as to ground states
providing, in principle, the calculation of the perturbation
corrections of large orders in the analytic or numerical form.
Besides this remarkable advantage over the standard approach to the
logarithmic perturbation theory, our method does not imply knowledge
of the exact solution for the zero approximation, which is obtained
automatically. And at last, the recursion formulae at hand, having
the same simple form both for the ground state and excited states,
can be easily adapted to applying any renormalization scheme for
improving the convergence of obtained series, as it is described
in~\cite{b32}.

This research was supported by a grant N 0106U000782 from the
Ministry of Education and Science of Ukraine which is gratefully
acknowledged.



\begin{thebibliography}{99}
\bibitem{b1} Sternheimer R 1951 {\it Phys. Rev.}
\textbf{84} 244
\bibitem{b2} Sternheimer R and Foley H 1956 {\it Phys. Rev.}
\textbf{102} 731
\bibitem{b3} Dalgarno A and Lewis J T 1955 {\it Proc. R. Soc.}
\textbf{A233} 70
\bibitem{b4} Schwartz C 1959 {\it Ann. Phys.} \textbf{2} 156
\bibitem{b5} Schwartz C and Tiemann J J 1959 {\it Ann. Phys.} \textbf{2}
178
\bibitem{b6} Mavromatis H A 1991 {\it Am. J. Phys.} \textbf{59}
738
\bibitem{b7} Zel'dovich Ya B 1956 {\it Zh. Eks. Teor. Fiz.}
\textbf{31} 1101
\bibitem{b8} Baz A I, Zel'dovich Ya B and Perelomov A M 1969
 \textit{Scattering, Reaction and Decay in Nonrelavistic Quantum Mechanics}
 (Jerusalem: Israel Program of Scientific Translation)
\bibitem{b9} Coutinho F A B, Nogami Y and Lauro Tomio 2000 {\it J. Phys. A: Math. Gen.} \textbf{33} 283
\bibitem{b10}  Ciffci H , Hall R L and Saad 2003 {\it J. Phys. A: Math. Gen.} \textbf{36}
11807
\bibitem{b11} Bayrak O and Boztosun I 2006 {\it J. Phys. A: Math. Gen.} \textbf{39}
6955
\bibitem{b12} Gonul B 2004 {\it Chin. Phys. Lett.} \textbf{21}
2330
\bibitem{b13} Gonul B, Celik N and Olgar E 2005 {\it Mod. Phys. Lett.} \textbf{A20}
1683
\bibitem{b14} Polikanov V S 1967 {\it Zh. Eks. Teor. Fiz.}
\textbf{52}  1326
\bibitem{b15} Polikanov V S 1975 {\it Teor. Mat. Fiz.}
\textbf{24} 230
\bibitem{b16} Dolgov A D and Popov V S 1978 {\it Phys. Lett.} \textbf{B79} 403
\bibitem{b17} Aharonov Y and Au C K 1979 {\it Phys. Rev.} \textbf{A20} 2245
\bibitem{b18} Turbiner A V 1984 {\it Usp. Fiz. Nauk} \textbf{144} 35
\bibitem{b19} Imbo T and Sukhatme U 1984 {\it Am. J. Phys.} \textbf{52} 140
\bibitem{b20} Rogers G M 1985 {\it J. Math. Phys.} \textbf{26} 567
\bibitem{b21} Mei W N and Chu D S 1998 {\it Phys. Rev.} \textbf{A58}
713
\bibitem{b22} Dobrovolska I V and Tutik R S 1999 {\it J. Phys.
A: Math. Gen.} \textbf{32} 563
\bibitem{b23} Zwaan A 1929 {\it Arch. Neerland. Sci. Exact. Natur. Ser.3} \textbf{A12} 1
\bibitem{b24} Dunham J L 1932 {\it Phys.Rev.} \textbf{41} 713
\bibitem{b25} Stepanov S S and Tutik R S 1991 {\it J. Phys. A: Math. Gen.} \textbf{24} L469
\bibitem{b26} Stepanov S S and Tutik R S 1991 {\it Zh. Eks. Teor. Fiz.} \textbf{100} 415
\bibitem{b27} Stepanov S S and Tutik R S 1992 {\it Teor. Mat. Fiz.} \textbf{90}
208
\bibitem{b28} Sergeenko M N 2000 E-print quant-ph/9912069
\bibitem{b29} Dolgov A D and Popov V S 1978 {\it Zh. Eks. Teor. Fiz.} \textbf{75}
2010
\bibitem{b30} Yukalova E P and Yukalov V I 1993 {\it J. Phys. A: Math. Gen.} \textbf{26}
2011
\bibitem{b31} Yukalov V I and Yukalova E P 1999 {\it Ann. Phys.} \textbf{277}
219
\bibitem{b32} Dobrovolska I V and Tutik R S 2001 {\it Int. J. Mod. Phys.} \textbf{A16} 2493
\bibitem{b33} Ikhdair S.M. and Sever Ramazan ArXiv:quant-ph/0603205.

 \end{thebibliography}
\end{document}